\documentclass[11pt]{article}
\usepackage{amssymb}
\usepackage{amsmath}
\usepackage{amscd}
\usepackage{latexsym}
\usepackage{bbm}

\catcode`@=11
\makeatletter\renewcommand{\section}{\@startsection
{section}{1}{\z@}{-3.5ex plus -1ex minus
    -.2ex}{2.3ex plus .2ex}{\large\bf }}

\makeatletter\renewcommand{\subsection}{\@startsection{subsection}{2}{\z@}{-3.25ex
plus -1ex minus
   -.2ex}{1.5ex plus .2ex}{\bf }}

\numberwithin{equation}{section}
\newcounter{saveeqn}

\catcode`@=12

\def\a{\alpha}
\def\b{\beta}
\def\ga{\gamma}
\def\la{\lambda}
\def\ga{\gamma}
\def\de{\delta}
\def\eps{\epsilon}
\def\ve{\varepsilon}
\def\vk{\varkappa}
\def\Si{\Sigma}

\def\vp{\varphi}
\def\th{\theta}
\def\O{\Omega}

\newcommand{\C}{\mathbb C}
\newcommand{\R}{\mathbb R}

\newcommand{\Gcal}{{\cal G}}
\newcommand{\Acal}{{\cal A}}

\newcommand{\Mcal}{{\cal M}}
\newcommand{\Fcal}{{\cal F}}
\newcommand{\Ncal}{{\cal N}}
\newcommand{\Tcal}{{\cal T}}

\newcommand{\gfrak}{{\mathfrak g}}

\newcommand{\ah}{{\hat{\smash{a}}}}

\newcommand{\hd}{{\hat{\textrm{d}}}}

\def\im{\textrm{i}}

\def\diff{\textrm{d}}
\def\pa{\mbox{$\partial$}}
\def\sfrac#1#2{{\textstyle\frac{#1}{#2}}}
\def\+{\dagger}
\def\={\ =\ }

\def\and{\quad\textrm{and}\quad}
\def\with{\quad\textrm{with}\quad}
\def\for{\quad\textrm{for}\quad}

\def\Id{\mathrm{Id}}

\begin{document}

\begin{titlepage}
\setcounter{page}{0}
\begin{flushright}
ITP-UH-11/16
\end{flushright}

\hspace{2.0cm}

\begin{center}

{\Large\bf Superstring theories as low-energy limit of\\[3pt] supergroup gauge theories }

\vspace{12mm}

{\large  Alexander D. Popov}\\[8mm]

\noindent {\em
Institut f\"ur Theoretische Physik \\
Leibniz Universit\"at Hannover \\
Appelstra\ss e 2, 30167 Hannover, Germany }\\
{Email: popov@itp.uni-hannover.de
}\\[6mm]

\vspace{10mm}

\begin{abstract}
\noindent We consider Yang-Mills theory with $N=2$ super translation group in $d=10$ auxiliary dimensions as the structure
group. The gauge theory is defined on a direct product manifold $\Sigma_2\times H^2$, where $\Sigma_2$ is a two-dimensional
Lorentzian manifold and $H^2$ is the open disc in $\R^2$ with the boundary $S^1=\partial H^2$. We show that in the adiabatic
limit, when the metric on  $H^2$ is scaled down, the Yang-Mills action supplemented by the $d=5$ Chern-Simons term becomes the
Green-Schwarz superstring action. More concretely, the Yang-Mills action in the infrared limit flows to the kinetic part of
the superstring action and the $d=5$ Chern-Simons action, defined on a 5-manifold with the boundary $\Sigma_2\times H^2$,
flows to the Wess-Zumino part of the superstring action. The same kind of duality between gauge fields and strings is
established for type IIB superstring on AdS$_5\times S^5$ background and a supergroup gauge theory with PSU(2,2$|$4) as the
structure group.
\end{abstract}

\end{center}
\end{titlepage}

\section {Introduction and summary}

\noindent {\bf Gauge/string correspondence: AdS/CFT  approach.} Superstring theory has a long history~\cite{1}-\cite{3} and
pretends on description of all four forces in Nature.  However, besides of all its well-known successes there is a number of
drawbacks, one of which is the lack of a good non-perturbative formulation. To overcome (partially) the last shortcoming it
was conjectured \cite{1t} that string theory in some backgrounds, including quantum gravity, is equivalent (dual) to a gauge
field theory. The most studied case is the AdS/CFT  correspondence \cite{1t, 2t, 3t} which conjectures that type IIB
superstring theory~\cite{12} in maximally supersymmetric AdS$_5\times S^5$ background, with $N$ units of self-dual five-form
flux, is equivalent to four-dimensional $\Ncal  = 4$ super-Yang-Mills theory with SU($N$) gauge group for large-$N$ limit and
fixed 't~Hooft parameter $\lambda =g^2_{YM} N$. In fact, for the AdS$_5\times S^5$ background and its zero-curvature (infinite
radius) limit $\R^{9,1}$ the Green-Schwarz superstring action has the form of a supercoset sigma model~\cite{12, 16, 4t}. On
classical level the AdS/CFT correspondence means a bijection between the moduli space $\Mcal_{string}$ of the $d=2$ sigma
model with the target space PSU(2,2$|$4)/SO(4,1)$\times$SO(5) and the moduli space $\Mcal_{gauge}$ of $\Ncal{=}4$
super-Yang-Mills theory in $d=4$ dimensions, defined on the boundary $\R^{3,1}$ of AdS$_5$. Note that the above sigma model is
coupled to gravity on $d=2$ worldsheet and the variation of the worldsheet metric\footnote{In ordinary sigma model the
worldsheet metric is fixed, it is not dynamical.} produces the Virasoro constraint equations which reflect the
reparametrization invariance imposed in string theory. Furthermore, superstring action should  also have so-called kappa
symmetry (fermionic) which dictates the form of an additional Wess-Zumino-type term~\cite{12, 16, 4t}.

\medskip

\noindent {\bf Gauge/sigma model correspondence: adiabatic approach.} For ordinary sigma models there is another
correspondence with Yang-Mills models based on the adiabatic approach and well studied in the literature. The adiabatic
approach to differential equations (based on introduction of ``slow" and ``fast" variables) exists more than 80 years and it
is used in many areas of physics. Briefly, one should multiply metric on the space parametrized by part of coordinates (they
will be ``fast") by $\ve^2$ with $\ve\in [0,1]$ and consider the limit $\ve\to 0$. In~\cite{5} the adiabatic limit method was
applied to the instanton equations defined on a direct product of two Riemannian surfaces $\Sigma_2$ and $\tilde \Sigma_2$
(with scaled down metric on $\tilde \Sigma_2$)  and it was shown that instanton solutions on $\Sigma_2\times\tilde \Sigma_2$
are in {\it one-to-one correspondence} with holomorphic maps  from $\Sigma_2$ into the moduli space $\Mcal_{\tilde \Sigma_2}$
of {\it flat connections} on $\tilde \Sigma_2$. Here, the Yang-Mills action reduces to the action of $d=2$ sigma model  on
$\Sigma_2$ with the target space $\Mcal_{\tilde \Sigma_2}$ while $\tilde \Sigma_2$ shrinks to a point. This result is closely
related with earlier observation by Atiyah~\cite{5t} that there is a {\it bijection} between the moduli space of
$G$-instantons\footnote{Here $G$ is the structure group of the gauge theory.} on the 4-sphere $S^4=\R^4\cup\{\infty\}$ and the
moduli space of {\it holomorphic maps} from $\C P^1\cong S^2$ into the based loop group $\Omega G= LG/G$. Here, the instanton
bundles are {\it framed} over $\{\infty\}\in S^4$, i.e. the gauge transformations are demanded to be identity at this point.
Later Donaldson assumed~\cite{8} that this result can be obtained via the adiabatic limit on the space $S^2\times H^2$ (which
is conformally diffeomorphic to $S^4\setminus S^1$), where $H^2$ is the shrinked 2-disc with the boundary $S^1$ and the loop
group $\Omega G$ appears as the map from $S^1=\partial  H^2$ into $G$. This assumption was proven in~\cite{9, 6} and it was
shown that the moduli space $\Mcal_\ve (G, k)$ of $d=4$ Yang-Mills $k$-instantons is bijective to the moduli space $\Mcal_0
(G, k)$ of sigma-model $k$-instantons for any $\ve\in [0,1]$ and any integer $k$.

Recall that in all above-mentioned cases the Yang-Mills action on the manifold $\Sigma_2\times\tilde \Sigma_2$ with the metric
$g^{}_{\Sigma_2} + \ve^2 g^{}_{\tilde\Sigma_2}$ reduces in the limit $\ve\to 0$ to the action of a sigma model on $\Sigma_2$
with the target space $\Mcal_{\tilde \Sigma_2}$ of flat connections  on $\tilde \Sigma_2$. The same result for the Lorentzian
signature with $\Sigma_2=\R^{1,1}$ and $\tilde \Sigma_2 = T^2$ (two-torus) was derived in~\cite{1a}. Furthermore,
in~\cite{1a}-\cite{4a} it was shown that $\Ncal =4$ super-Yang-Mills theory on $\Sigma_2\times\tilde \Sigma_2$ flows to a
supersymmetric  or topological sigma model on $\Sigma_2$ while $\tilde \Sigma_2$ shrinks to a point. The fact that the
Yang-Mills action reduces for $\ve\to 0$ to a sigma-model action, which describes also non-instantonic solutions, leads to a
reasonable assumption~\cite{6t, 7t} that the moduli space of solutions of the full Yang-Mills equations on $\R^{4,0}$ is
bijective to the moduli space of {\it harmonic maps} from $S^2=\R^2\cup\{\infty\}$ to $\Omega G$. For the Minkowski space
$\R^{3,1}$ this is also so, as was demonstrated in~\cite{8t}.

\medskip

\noindent {\bf Gauge/string correspondence via the adiabatic approach.} Superstring sigma models differ from ordinary sigma models by coupling
with gravity and by using superspaces instead of ordinary manifolds as target spaces.\footnote{In this paper we restrict ourselves
to the supercoset PSU(2,2$|$4)/SO(4,1)$\times$SO(5) and its ``flat" limit.} Therefore, to rise the adiabatic correspondence between Yang-Mills
models and sigma models to the Green-Schwarz superstring case one should

i) couple Yang-Mills to gravity,

ii) consider supergroups instead of ordinary Lie groups as the structure groups in gauge theory.

\smallskip

For the case of Yang-Mills models with Higgs fields the task i) was accomplished by Shifman and Yung~\cite{SY}.\footnote{For
further study of Yang-Mills-Higgs/string correspondence see the papers~\cite{KSY, KSYS}.} More precisely, $\Ncal =2$
supersymmetric U(2) Yang-Mills-Higgs theory on $\Sigma_2\times \R^2$ with a Fayet-Illiopoulos term and four flavour
hypermultiplets in the fundamental of U(2) admits semilocal vortices on $\R^2$  whose moduli are parametrized by
$X^{10}=\R^4\times Y^6$, where $Y^6$ is the resolved conifold. Allowing the vortex moduli to depend on the coordinates of
$\Sigma_2$ yields a {\it string sigma model} with worldsheet $\Sigma_2$ and target $X^{10}$, which describes the effective
vortex dynamics.\footnote{Unfortunately, I learned about these interesting results only in June 2016 that did not give me the
opportunity to quote them earlier.}

For pure Yang-Mills theory, the task i) was accomplished in~\cite{7}, where a correspondence between a Yang-Mills model on
$\Sigma_2\times H^2$ and bosonic strings with a worldsheet $\Sigma_2$ was established. Coupling of Yang-Mills to gravity leads
to vanishing of Yang-Mills energy-momentum tensor along $\Sigma_2$ and in the adiabatic limit $\ve\to 0$ these components of
energy-momentum tensor give the Virasoro constraint equations of bosonic strings. These constraints may be responsible for
restoring of unitarity both on string and Yang-Mills levels. In a conformally flat gauge $\Sigma_2\to \R^{1,1}$ the action
reduces to the usual $d=2$ Minkowski space form but it is supplemented by the Virasoro constraints.

In this paper we will consider the step ii) and derive a correspondence between the Green-Schwarz superstring theory (of type
I, IIA and IIB) and $d=4$ supergroup gauge theory on $\Sigma_2\times H^2$ by using the adiabatic approach. Note that
supergroup gauge theories were considered in the literature (see e.g.~\cite{MW, 12t} and references therein). Their quantization
is not well studied yet and our discussion will therefore be purely classical. Due to deep relations of supergroup gauge
theories with superstring theories, as was discussed in~\cite{12t} and will be shown below, they definitely deserve more deep
study. In this paper we will consider the simplest case of the supertranslation group $G$ in ten auxiliary dimensions as the
gauge group and only briefly discuss the case of the supergroup PSU(2,2$|$4) related with AdS$_5\times S^5$ since it is
treated similarly. Our discussion will be close to that for the bosonic string~\cite{7}. This is natural since in both cases
the constructions are based on the adiabatic approach. However, we will provide here more explanations for each step,
especially in description of moduli space of flat connections and related superstring target space. In particular, it will be
shown that stringy $d=2$ Wess-Zumino term appears from the $d=5$ Chern-Simons term.

The organization of this paper is as follows. In Section 2 we describe a 4-manifold $M^4=\Sigma_2\times H^2$, the
$\ve$-deformed metric on it and introduce $\ve$-dependent Yang-Mills action on $M^4$ with a gauge supergroup $G$. In Section 3
we describe the adiabatic limit  $\ve\to 0$ under which the Yang-Mills theory reduces to a stringy sigma model. Besides
reduction of gauge field equations and action, we describe in detail how gauge field moduli become coordinates on sigma-model
target space. Effective action and Virasoro-type constraints will be derived. In Section 4 we describe how $B$-field and
Wess-Zumino-type terms can be obtained from Yang-Mills models and briefly comment on superstrings in AdS$_5\times S^5$
background. In Conclusions we summarize our findings and make some further remarks concerning some issues that can be done in
future work.

\section{Supergroup Yang-Mills theory}

\noindent {\bf Lie supergroup $G$.}  We consider Yang-Mills theory on a direct product manifold $M^4=\Si_2\times H^2$,
where $\Si_2$ is a two-dimensional  Lorentzian manifold (flat case is included) with local coordinates $x^a, a,b,...=1,2$, and
a metric tensor $g^{}_{\Si_2}=(g_{ab})$, $H^2$ is the open disc with coordinates $x^i$, $i,j,...= 3,4$ and the metric tensor
$g^{}_{H^2}=(g_{ij})$. Then $(x^\mu )=(x^a, x^i)$ are local coordinates on $M^4$ with metric tensor $(g_{\mu\nu})= (g_{ab},
g_{ij} ), \mu , \nu = 1,...,4$.

As the structure group of Yang-Mills theory, we consider  the coset $G{=}$SUSY$(N{=}2)/SO(9,1)$ (cf. \cite{16}) which is the
subgroup of $N{=2}$ super Poincare group in ten auxiliary dimensions generated by translations and $N{=}2$ supersymmetry
transformations. Its generators $(\xi_{\a}, \xi_{Ap})$ obey the Lie superalgebra $\gfrak =\,$Lie$\,G$,
\begin{equation}\label{1}
\{\xi_{Ap}, \xi_{Bq}\} =(\ga^\a C)_{AB}\de_{pq}\xi_\a\ ,\quad [\xi_{\a}, \xi_{Ap}]=0\ ,\quad [\xi_{\a}, \xi_{\b}]=0\ ,
\end{equation}
where $\ga^\a$ are the $\ga$-matrices, $C$ is the charge conjugation matrix, $\a = 0,...,9,\ A=1,...,32$ and $p,q=1,2$ label
the number of supersymmetries.  Coordinates on $G$ are $X^\a$ and two spinors $\th^{Ap}$ of the
Majorana-Weyl type. On the superalgebra $\gfrak=\,$Lie$\,G$ we introduce the metric $\langle\cdot \rangle$ with components
\begin{equation}\label{2}
\langle\xi_{\a}\, \xi_{\b}\rangle=\eta_{\a\b}\ , \quad \langle\xi_{\a}\, \xi_{Ap}\rangle=0\and \langle\xi_{Ap}\, \xi_{Bq}\rangle=0\ ,
\end{equation}
where $(\eta_{\a\b})=\,$diag$(-1,1,...,1)$ is the Lorentzian metric on $\R^{9,1}$ and the last equality in (\ref{2}) is
standard in superstring theory.

\medskip

\noindent {\bf Yang-Mills action.} We consider the gauge potential $\Acal =\Acal_{\mu}\diff x^\mu$ with values in $\gfrak$
and the $\gfrak$-valued  gauge field
\begin{equation}\label{3}
 \Fcal =\sfrac12\Fcal_{\mu\nu}\diff x^\mu \wedge \diff x^\nu\with \Fcal_{\mu\nu} =\partial_\mu\Acal_\nu - \partial_\nu\Acal_\mu
 + [\Acal_\mu , \Acal_\nu]\ ,
\end{equation}
where $[\ ,\ ]$ is the commutator or anti-commutator for two $\xi_{Ap}$-generators. On  $M^4=\Si_2\times H^2$ we have the
obvious splitting
\begin{equation}\label{4}
 \diff s^2 = g_{\mu\nu}\diff x^\mu \diff x^\nu = g_{ab}\diff x^a \diff x^b + g_{ij}\diff x^i \diff x^j\ ,
\end{equation}
\begin{equation}\label{5}
\Acal =\Acal_{\mu}\diff x^\mu= \Acal_{a}\diff x^a+\Acal_{i}\diff x^i\ ,
\end{equation}
\begin{equation}\label{6}
 \Fcal =\sfrac12\Fcal_{\mu\nu}\diff x^\mu \wedge \diff x^\nu =\sfrac12\Fcal_{ab}\diff x^a \wedge \diff x^b + \Fcal_{ai}\diff x^a \wedge \diff x^i
+\sfrac12\Fcal_{ij}\diff x^i \wedge \diff x^j\ .
 \end{equation}

By using the adiabatic approach in the form presented in \cite{5, 6}, we deform the metric (\ref{4}) and introduce
\begin{equation}\label{7}
 \diff s^2_\ve = g_{\mu\nu}^\ve\,\diff x^\mu \diff x^\nu = g_{ab}\diff x^a \diff x^b + \ve^2g_{ij}\diff x^i \diff x^j\ ,
\end{equation}
where $\ve\in [0,1]$ is a real parameter. Then $\det (g_{\mu\nu}^\ve )=\ve^4\det (g_{ab}) \det (g_{ij})$ and
\begin{equation}\label{8}
 \Fcal^{ab}_\ve = g_\ve^{ac}g_\ve^{bd}\Fcal_{cd}= \Fcal^{ab}\ ,\quad \Fcal^{ai}_\ve = g_\ve^{ac}g_\ve^{ij}\Fcal_{cj}=
 \ve^{-2}\Fcal^{ai}\and
 \Fcal^{ij}_\ve = g_\ve^{ik}g_\ve^{jl}\Fcal_{kl}=\ve^{-4}\Fcal^{ij}\ ,
\end{equation}
where indices in $\Fcal^{\mu\nu}$ are raised by the non-deformed metric tensor $g^{\mu\nu}$. It is assumed that  $\Fcal^{\mu\nu}$
smoothly depend on $\ve$ with well-defined limit for $\ve\to 0$.

For the deformed metric  (\ref{7}) the Yang-Mills action functional is
\begin{equation}\label{9}
S_\ve=\frac{1}{2\pi}\,\int_{M^4} \diff^4x\,\sqrt{|\det g^{}_{\Si_2}|}\,\sqrt{\det g^{}_{H^2}}\,\left\{\ve^2\langle\Fcal_{ab}\,
\Fcal^{ab}\rangle + 2\langle\Fcal_{ai}\, \Fcal^{ai}\rangle + \ve^{-2}\langle\Fcal_{ij}\, \Fcal^{ij}\rangle\right\}\ ,
\end{equation}
where $\pi$ is the ``area" of the disc $H^2$ of radius $R=1$.

\medskip

\noindent {\bf Remark.} On the disc $H^2$ of radius $R=1$ one can consider both the flat metric $g_{ij}=\de_{ij}$ (then
${\mbox{Vol}(H^2)}=\pi$) and the metric
\begin{equation}\label{10}
g_{ij}=\frac{4}{(1-r^2)^2}\,\de_{ij}\with r^2=\de_{ij}x^ix^j\ .
\end{equation}
However, we will see later that in all integrals over $H^2$ the metric $g_{ij}$ enters in the combination $\sqrt{\det
g^{}_{H^2}}\,g^{ij}\xi_i\xi_j=\de^{ij}\xi_i\xi_j$ for one-forms $\xi=\xi_i\diff x^i$ on $H^2$. Hence all calculations for the
metric  (\ref{10}) are equivalent to the calculations  for $g_{ij}=\de_{ij}$. That is why we will consider the flat metric on
$H^2$ as in many mathematical papers considering Yang-Mills theory on the $n$-balls with $n\ge 2$.

\medskip

\noindent {\bf Coupling to gravity.} In the Introduction we mentioned that our Yang-Mills model should be coupled to gravity.
In the considered case we should add to the Lagrangian (\ref{9}) the term
\begin{equation}\label{11}
\sqrt{|\det g^{\ve}_{M^4}|}\, R^{\ve}_{M^4} =\sqrt{|\det g^{}_{\Si_2}|}\,\sqrt{\det g^{}_{H^2}}\,\left\{\ve^2 R_{\Si_2} +
\ve^2 R^{\ve}_{H^2}\right\}\ ,
\end{equation}
where $R^{\ve}_{M^4}$ is the scalar curvature of the manifold $M^4$ with the metric (\ref{7}), $R_{\Si_2}$ is the scalar
curvature of $\Si_2$ and $R^{\ve}_{H^2}$ is the scalar curvature of $H^2$ with the metric multiplied by $\ve^2$. For any
choice of metrics on $\Si_2$ and $H^2$ the terms in the bracket in (\ref{11}) are constants and these terms do not contribute
to the equations of motion. For that reason we do not add them to the Lagrangian (\ref{9}). For simplicity we also do not
consider coupling to dilaton field.

\medskip

\noindent {\bf Field equations.}
For the deformed metric (\ref{7}) the Yang-Mills equations have the form
\begin{equation}\label{12}
\ve^2 D_a\Fcal^{ab} + D_i \Fcal^{ib} =0\ ,
\end{equation}
\begin{equation}\label{13}
D_a\Fcal^{aj} + \ve^{-2}D_i \Fcal^{ij} =0\ ,
\end{equation}
where $D_a$, $D_i$ are Yang-Mills covariant derivatives on the curved background $M^4=\Si_2\times H^2$.

The metric on $H^2$ is fixed but it is not fixed on $\Si_2$ and the Euler-Lagrange equations for $g^{}_{\Si_2}$ yield the constraint equations
(remnant of Einstein-Yang-Mills equations)
\begin{equation}\label{14}
 T_{ab}^{\ve}{=}\ve^2\left\{g^{cd}\langle\Fcal_{ac}\,\Fcal_{bd}\rangle{-}\sfrac14 g_{ab}\langle\Fcal_{cd}\,\Fcal^{cd}\rangle\right\}{+}g^{ij}
\langle\Fcal_{ai}\, \Fcal_{bj}\rangle{-}\sfrac12
 g_{ab}\langle\Fcal_{ci}\, \Fcal^{ci}\rangle{-}\sfrac14 \ve^{-2}g_{ab}\langle\Fcal_{ij}\,\Fcal^{ij}\rangle{=}0
\end{equation}
for the Yang-Mills energy-momentum tensor $T_{\mu\nu}^{\ve}$, i.e. its components along $\Si_2$ are vanishing. Other components of  $T_{\mu\nu}^{\ve}$,
besides (\ref{14}), are not constrained.

\section{Low-energy effective action}

\noindent {\bf Adiabatic limit.} The term $\ve^{-2}\langle\Fcal_{ij}\, \Fcal^{ij}\rangle$ in the Yang-Mills action
(\ref{9}) diverges when
 $\ve\to 0$. To avoid this we impose the flatness condition
\begin{equation}\label{15}
 \Fcal_{ij}=0
\end{equation}
on the components of the field tensor along $H^2$ for $\ve =0$. However, for $\ve>0$ the condition (\ref{15}) is not needed
and one can consider $ \Fcal_{ij} (\ve>0)\ne 0$, only $\Fcal_{ij}(\ve =0)= 0$.

In the adiabatic limit $\ve\to 0$, the Yang-Mills action (\ref{9}) becomes
\begin{equation}\label{16}
S_0=\frac{1}{\pi}\int_{M^4} \diff^4x\,\sqrt{|\det g^{}_{\Si_2}|}\,\langle\Fcal_{ai}\, \Fcal^{ai}\rangle
\end{equation}
with the equations of motion
\begin{equation}\label{17}
D_i\Fcal^{ib}:=\frac{1}{\sqrt{|\det g^{}_{\Si_2}|}}\,\partial_i\left(\sqrt{|\det
g^{}_{\Si_2}|}\,\de^{ij}g^{ab}\Fcal_{aj}\right) + [\Acal_i , \Fcal^{ib}]=0\ ,
 \end{equation}
\begin{equation}\label{18}
D_a\Fcal^{aj}:=\frac{1}{\sqrt{|\det g^{}_{\Si_2}|}}\,\partial_a\left(\sqrt{|\det
g^{}_{\Si_2}|}\,\de^{ij}g^{ab}\Fcal_{ib}\right) + [\Acal_a , \Fcal^{aj}]=0\ .
 \end{equation}

One can see that the constraint equations (\ref{14})  are non-singular in the limit $\ve\to 0$ due to (\ref{15}):
\begin{equation}\label{19}
T^0_{ab}=\de^{ij} \langle\Fcal_{ai}\, \Fcal_{bj}\rangle -\sfrac12
 g_{ab}\langle\Fcal_{ci}\, \Fcal^{ci}\rangle =0\ .
\end{equation}
They also follow from (\ref{16}) as Euler-Lagrange equations for  $g^{}_{\Si_2}$. In general, for $\ve\in [0,1]$ we assume
that fields $\Acal_\mu$ and $\Fcal_{\mu\nu}$ smoothly depend on $\ve$ and can be expanded in power series in $\ve$, e.g.
$\Acal_\mu=\Acal_\mu^0+\ve^3\Acal_\mu^1+...\ $. Note that $\Fcal^{\mu\nu}(\ve)$ should not be confused with
$\Fcal^{\mu\nu}_\ve = g_\ve^{\mu\sigma}g_\ve^{\nu\la}\Fcal_{\mu\nu}(\ve)$ in (\ref{8}). We omit $\ve$ from
$\Fcal_{\mu\nu}(\ve)$ for simplicity of notation. In (\ref{15})-(\ref{19}) we have zero terms in $\ve$ and omit index ``0"
from the fields. In fact, in (\ref{15}) we have $\Fcal_{ij}^0 =\partial_i\Acal_j^0 -\partial_j\Acal_i^0
 + [\Acal_i^0 , \Acal_j^0]=0$ but  $\Fcal_{ij}^1$, $\Fcal_{ij}^2$ etc. must not be zero.

\medskip

\noindent {\bf Flat connections. } Consider first the adiabatic  flatness equation (\ref{15}). Flat connection
$\Acal_{H^2} := \Acal_i\diff x^i = \Acal_i(\ve{=}0)\diff x^i$ on $H^2$ has the form
\begin{equation}\label{20}
 \Acal_{H^2}=g^{-1}\hd g\with \hd =\diff x^i\pa_i\for\pa_i=\frac{\pa}{\pa x^i}\ ,
\end{equation}
where $g$ is a smooth map from $H^2$ into the gauge supergroup $G$ for any fixed $x^a\in\Si_2$. We denote by $C^\infty (H^2,
G)$ the space of flat connections on $H^2$ given by (\ref{20}). On $H^2$, as on a manifold with boundary, the (super)group of
gauge transformations is defined as (see e.g. \cite{8, 9, 6})
\begin{equation}\label{21}
 \Gcal_{H^2}= \left\{g: H^2\to G\mid g^{}_{|\pa H^2}=\Id\right\}\ .
\end{equation}
Hence the solution space of the equation (\ref{15})  is the infinite-dimensional space $C^\infty (H^2, G)$ and the moduli
space  is the loop (super)group
\begin{equation}\label{22}
 \Mcal = C^\infty(H^2, G)/  \Gcal_{H^2}= LG\ .
\end{equation}
Here $LG=C^\infty(S^1, G)$ is the loop supergroup of smooth maps from the circle
$S^1=\pa H^2$ into $G$.

\medskip

\noindent {\bf Moduli space. } From (\ref{20})-(\ref{22}) it is clear that $\Acal^{}_{H^2}$ is defined by its value
$\Acal^{}_{S^1}$ on the boundary $S^1=\partial H^2$ of the disc $H^2$, parametrized by $\exp(2\pi\im\vp)\in S^1$ with $\vp\in
[0,1]$. The connection $\Acal^{}_{S^1}$ belongs to the space $\O^1(S^1, \gfrak )$ of one-forms on $S^1$ with values in $\gfrak
=\,$Lie$\,G$. It is the affine space of connections on the trivial bundle $S^1\times G$ over $S^1$ and the loop group $LG$
acts on $\O^1(S^1, \gfrak )$ by transformations
\begin{equation}\label{23}
f\in LG : \Acal^{}_{S^1}\mapsto \Acal^{f}_{S^1}= f \Acal^{}_{S^1} f^{-1} + f\breve\diff f^{-1} \with
\breve\diff=\diff\vp\pa_\vp\ .
\end{equation}

For any $\vp\in [0,1]$ we introduce the map
\begin{equation}\label{24}
h_\vp : \O^1(S^1, \gfrak )\to G
\end{equation}
which is defined as the unique solution of the differential equation \cite{10t}
\begin{equation}\label{25}
h^{-1}_\vp(\Acal^{}_{S^1})\,\breve\diff h_\vp(\Acal^{}_{S^1})= \Acal^{}_{S^1}\ ,\quad h_0(\Acal^{}_{S^1})=\Id\ .
\end{equation}
For this map we have the equivariance condition~\cite{10t}
\begin{equation}\label{26}
h_\vp(\Acal^{f}_{S^1})=f(0)h_\vp (\Acal^{}_{S^1}) f^{-1}(\vp)\ ,
\end{equation}
where $\Acal^{f}_{S^1}$ is given in (\ref{23}). Note that $h_\vp$ is not periodic in $\vp$, i.e. $h_1\ne h_0$.
In fact, $h_1$ is the Wilson loop defining the holonomy of $\Acal^{}_{S^1}$.

Recall that the based loop group $\O G\subset LG$ is defined as the kernel of the evaluation mapping
$LG\to G$, $f(\vp )\mapsto f(1)$. Let us introduce the {\it holonomy map}
\begin{equation}\label{27}
h_1 : \O^1(S^1, \gfrak )\to G\ ,
\end{equation}
where $h_\vp$ is defined by (\ref{25})~\cite{10t}. Then from (\ref{25}) and (\ref{26}) one sees that the action of
$\O G$ on $\O^1(S^1, \gfrak )$ is free,
\begin{equation}\label{28}
h_1(\Acal^{f}_{S^1})= f(0)h_1(\Acal^{}_{S^1})\quad\Leftrightarrow\quad \Acal^{f}_{S^1}(1)=\Acal^{}_{S^1}(1)\for f\in\O G
\end{equation}
and the quotient of  $\O^1(S^1, \gfrak )$ by $\O G$ is just the holonomy map (\ref{27}). Thus, (\ref{27}) is the projection
in the $\O G$-principal bundle over $G$. Hence, if we choose all $g$ in (\ref{20}) such that
\begin{equation}\label{29}
g^{}_{|S^1}=h_\vp
\end{equation}
with $h_\vp\in G$ given by (\ref{25}), then we restrict ourselves to the subspace $G\subset LG$ in the moduli space $LG$.
Since this restriction is invariant with respect to $\O G=LG/G$, it is consistent with the equations of
motion.\footnote{Simply put, the restriction of $g$ from (\ref{20}) to $S^1=\pa H^2$ is an arbitrary element $g(\vp)$ of the
loop group $LG$. It gives arbitrary gauge potential on $S^1$, $\Acal^{}_{S^1} = g^{-1}(\vp)\breve\diff g(\vp)\in \O^1(S^1,
\gfrak )$. Then we choose one element $g(\vp)$ from $C^\infty (S^1,G)$ (with fixed dependence on $\vp$) such that $g(0)=\Id$,
$g(1)\ne \Id$ and denote it by $h_\vp$. Multiplying $h_\vp$ by arbitrary element $\tilde g$ of the group (\ref{21}), we get
$g=\tilde g h_\vp\in C^\infty (H^2,G)$ with the moduli space $G\subset LG$.} Note that $\O G$ is not a subgroup in the gauge
group (\ref{21}) whose elements are identity on $S^1=\pa H^2$. Thus, from now our moduli space is the Lie supergroup $G\subset
LG$ on which we have coordinates $(X^a, \th^{Ap})$ introduced in Section 2.

In the adiabatic approach it is assumed that $\Acal_\mu = \Acal_\mu (x^a, x^i, X^\a , \th^{Ap})$ depend on
$x^a\in \Si_2$ only via moduli parameters \cite{10,11}, i.e. $\Acal_\mu = \Acal_\mu (X^\a (x^a), \th^{Ap}(x^a), x^i)$.
Then moduli of gauge fields define the map
\begin{equation}\label{30}
(X, \th ): \Si_2\to G \with  (X (x^a), \th (x^a))= \left\{X^\a (x^a), \th^{Ap}(x^a)\right\}\ ,
\end{equation}
where $G$ is now our moduli space. Acting by gauge transformations from (\ref{21}) on flat connections $\Acal_i$ in (\ref{20})
which depend only on moduli $(X, \th )$ defined by (\ref{25})-(\ref{29}), we obtain the subspace $\Ncal$ in the full solution
space $C^\infty(H^2, G)$. The moduli space of these solutions is
\begin{equation}\label{31}
 G=\Ncal /\Gcal\ ,
\end{equation}
where $\Gcal = \Gcal^{}_{H^2}$ for any fixed $x^a\in \Si_2$.

The maps (\ref{30}) are constrained by the equations (\ref{17})-(\ref{19}). Since $\Acal^{}_{H^2}$ is a flat connection for any
$x^a\in \Si_2$, the derivatives  $\pa_a\Acal_i$ have to satisfy the linearized (around  $\Acal^{}_{H^2}$) flatness condition,
i.e.  $\pa_a\Acal_i$ belong to the tangent space $\Tcal_\Acal\Ncal$ of the space $\Ncal$. Using the projection $\pi : \Ncal\to G$
with fibres $\Gcal$,
one can decompose $\pa_a\Acal_i$ into the two parts
\begin{equation}\label{32}
T_\Acal\Ncal= \pi^*T_\Acal G\oplus T_\Acal \Gcal \quad\Leftrightarrow\quad\pa_a\Acal_i=\Pi_a^\a\xi_{\a i} +
(\pa_a\th^{Ap})\xi_{Api} + D_i\eps_a \ ,
\end{equation}
where
\begin{equation}\label{33}
\Pi_a^\a :=\pa_a X^\a - \im \de_{pq}\bar\th^p\ga^\a\pa_a\th^q\ ,
\end{equation}
$\eps_a$ are $\gfrak$-valued gauge parameters ($\diff x^iD_i\eps_a\in T_\Acal \Gcal$) and  $\{\xi_{\a}=\xi_{\a i}\diff x^i ,
\xi_{Ap}=\xi_{Ap i}\diff x^i\}\in T_\Acal G$ can be identified with $\gfrak =\,$Lie$\,G$.

The gauge parameters $\eps_a$ are determined by the gauge fixing conditions
\begin{equation}\label{34}
  \de^{ij} D_i\xi^{}_{\Delta j}=0\quad\Rightarrow\quad \de^{ij} D_iD_j\eps_a=\de^{ij} D_i \pa_a\Acal_j \ ,
\end{equation}
where the index $\Delta$ means $\a$ or $Ap$. It is easy to see that
\begin{equation}\label{35}
  \de^{ij} D_i\xi_{\a j}=0\quad\Rightarrow\quad  \xi_{\a i}=\ve_{ij}D_j\xi_\a = \ve_{ij}\pa_j\xi_\a
\end{equation}
and similarly $\xi_{Ap i}=\ve_{ij}D_j\xi_{Ap}$ since $\Fcal_{ij}=[D_i, D_j]=0$\ .

\medskip

\noindent {\bf Effective action. } Recall that $\Acal_i$ are given by (\ref{20}) and $\Acal_a$ are yet free. In the
adiabatic approach one choose $\Acal_a=\eps_a$ \cite{10, 11} and $\eps_a$ are defined from (\ref{34}). Then we obtain
\begin{equation}\label{36}
 \Fcal_{ai}=\pa_a\Acal_i - D_i\Acal_a = \Pi_a^\b\xi_{\b i} + (\pa_a\th^{Ap})\xi_{Api}=:\Pi^{\Delta}_a\xi_{\Delta i} \in T_\Acal G\ .
\end{equation}
Substituting  (\ref{36}) into (\ref{17}), we see that  (\ref{17}) is resolved due to  (\ref{34}). Substituting  (\ref{36}) in
(\ref{18}), we will get the equations of motion for $X^\a (x^a), \th^{Ap}(x^a)$ following from the action (\ref{16}) which
after inserting  (\ref{36}) into  (\ref{16})  and integrating over $H^2$ become
\begin{equation}\label{37}
S_0=\int_{\Si_2} \diff x^1 \diff x^2\,\sqrt{|\det g^{}_{\Si_2}|}\, g^{ab}\,\Pi_a^\a\,\Pi_b^\b\,\eta_{\a\b}\ .
\end{equation}
This is the kinetic part of the Green-Schwarz superstring action. Note that
\begin{equation}\label{38}
 \eta_{\a\b}=\frac{1}{\pi} \int \diff x^3 \diff x^4\,\langle\xi_{\a i}\,\xi_{\b j}\rangle \de^{ij}=
\frac{1}{\pi}\int  \diff x^3 \diff x^4\,\langle\pa_i\xi_{\a }\,\pa_j\xi_{\b }\rangle \de^{ij}\ ,
\end{equation}
since vector fields $\xi_{\Delta}$ on $G$ can be identified with the generators of the Lie algebra Lie$\,G$ described in
(\ref{1}), (\ref{2}). As we mentioned in Section 2, this result does not depend on which metric ($g_{ij}=\de_{ij}$ or $g_{ij}$
from  (\ref{10})) we choose on the disc $H^2$. Substituting  (\ref{36}) into the constraint equations  (\ref{19}) and
integrating them over $H^2$, we obtain the equations
\begin{equation}\label{39}
 \eta_{\a\b} \,\Pi_a^\a\,\Pi_b^\b - \sfrac12\, g_{ab}\, g^{cd}\,\eta_{\a\b} \,\Pi_c^\a\,\Pi_d^\b =0\ ,
\end{equation}
which can also be derived from  (\ref{37}) by variation of the metric $g^{ab}\mapsto\de g^{ab}$.

\section{Topological terms}

\noindent {\bf Wess-Zumino-type term. } The action  (\ref{37}) is not yet the full Green-Schwarz action which contains
additional Wess-Zumino-type term \cite{12}.
This term is described as follows. One considers a Lorentzian 3-manifold $\Si_3$
with the boundary $\pa\Si_3=\Si_2$ and coordinates $x^\ah, \ah =0,1,2$. On  $\Si_3$  one introduces the 3-form \cite{16}
\begin{equation}\label{40}
\O_3= \im\,\diff x^\ah\Pi_\ah^\a\wedge(\check\diff\bar\th^1\ga^\b\wedge\check\diff\th^1 -
\check\diff\bar\th^2\ga^\b\wedge\check\diff\th^2) \,\eta_{\a\b}=\check\diff\O_2\ \with \check\diff= \diff x^\ah\frac{\pa}{\pa x^\ah},
\end{equation}
where
\begin{equation}\label{41}
\O_2=-\im\diff X^\a\wedge(\bar\th^1\ga^\b\diff\th^1 -
\bar\th^2\ga^\b\diff\th^2)\eta_{\a\b} + \bar\th^1\ga^\a\diff\th^1 \wedge
\bar\th^2\ga^\b\diff\th^2\eta_{\a\b}\with \diff= \diff x^a\frac{\pa}{\pa x^a}\ .
\end{equation}
Then the term
\begin{equation}\label{42}
 S_{WZ} = \int_{\Si_3}\O_3 = \int_{\Si_2}\O_2
\end{equation}
is added to the functional (\ref{37}) and the Green-Schwarz action is
\begin{equation}\label{43}
  S_{GS}= S_{0} + \vk S_{WZ}
\end{equation}
with a properly chosen real coefficient $\vk$.

We look for a topological type addition to the Yang-Mills action (\ref{9}) which in the infrared limit $\ve\to 0$ will give us
this Wess-Zumino-type term. We will show that this is the standard Chern-Simons term in $d=5$ dimensions equivalent to the 3rd
Chern character in $d=6$. To show this we extend $H^2$ to a Riemannian 3-manifold $B^3$ with the boundary $\pa B^3= H^2$ and
coordinates $x^{\hat \imath}$, $\hat\imath =3,4,5$. Also, we extend $\Si_2$ to $\Si_3$ described above and consider the
6-manifold $M^6=\Si_3\times B^3$ with coordinates $(x^{\hat\mu})=(x^{\hat a}, x^{\hat \imath})$. Note that in addition to the
components $\Fcal_{ai}$ in (\ref{36}) of Yang-Mills fields we now have the components $\Fcal_{0i}, \Fcal_{05}, \Fcal_{a5}$ and
in the limit when $B^3$  shrinks to a point all these components have the form
\begin{equation}\label{43a}
\hat\Fcal_{\hat a\hat\imath}=(\pa_{\hat a} X^\a - \im \de_{pq}\bar\th^p\gamma^\a\pa_{\hat a} \th^q)\xi_{\a\hat\imath} +
(\pa_{\hat a}\th^{Ap})\xi_{Ap\hat\imath}
=:\hat\Pi_{\hat a}^\Delta\xi_{\Delta\hat\imath}\ ,
\end{equation}
where $\xi_{\Delta\hat\imath}=\ve_{\hat\imath\hat\jmath}D_{\hat\jmath}\xi_\Delta$ and the index $\Delta$ means $\a$ or $Ap$,
as discussed earlier.

Let us consider on $M^6=\Si_3\times B^3$ the gauge field $\hat\Fcal =\sfrac12\hat\Fcal_{\hat\mu\hat\nu} \diff
x^{\hat\mu}\wedge\diff x^{\hat\nu}$ and the topological Yang-Mills term
\begin{equation}\label{43b}
S_{top}=\int_{\Si_3\times B^3}\langle \hat\Fcal\wedge\hat\Fcal\wedge\hat\Fcal\rangle\ ,
\end{equation}
where $\langle \hat\Fcal\wedge\hat\Fcal\wedge\hat\Fcal\rangle = \diff_{M^6}CS_5(\hat\Acal )$ is the 3rd Chern character and
$CS_5(\hat\Acal )=\langle \hat\Acal\wedge\hat\Fcal\wedge\hat\Fcal\rangle -\sfrac12 \langle
\hat\Acal\wedge\hat\Acal\wedge\hat\Acal\wedge\hat\Fcal\rangle + \sfrac{1}{10}\langle
\hat\Acal\wedge\hat\Acal\wedge\hat\Acal\wedge\hat\Acal\wedge\hat\Acal\rangle$ is the Chern-Simons 5-form. The manifold
$\Si_3\times B^3$ has the boundary $\Si_2\times B^3\cup \Si_3\times H^2$ and the term (\ref{43b}) can be rewritten as the
integral of $CS_5(\hat\Acal )$ over the boundary of $\Si_3\times B^3$. Note that $\Si_2\times H^2$ is the common boundary of
$\Si_2\times B^3$ and $\Si_3\times H^2$ and, as usual, one takes $\hat\Acal_0=0=\hat\Acal_5$ on $\Si_2\times H^2$ as the
boundary condition for the Chern-Simons actions.

In the adiabatic limit $\ve\to 0$ the functional (\ref{43b}) becomes
\begin{equation}\label{43c}
S_{top}=\int_{\Si_3\times B^3} \hat\Pi^\Gamma\wedge\hat\Pi^\Delta\wedge\hat\Pi^\Lambda\wedge\langle
\tilde\diff\xi_\Gamma\wedge\tilde\diff\xi_\Delta\wedge\tilde\diff\xi_\Lambda\rangle
=\int_{\Si_3}f_{\Gamma\Delta\Lambda}\hat\Pi^\Gamma\wedge\hat\Pi^\Delta\wedge\hat\Pi^\Lambda = \int_{\Si_3}\O_3 =
\int_{\Si_2}\O_2\ ,
\end{equation}
where $\hat\Pi^\Gamma = \hat\Pi^\Gamma_{\hat a}\diff x^{\hat a}$, $\O_3$ and $\O_2$ are the forms given by (\ref{40}),
(\ref{41}) and
\begin{equation}\label{43e}
f_{\Gamma\Delta\Lambda} = \int_{B^3}\langle \tilde\diff\xi_\Gamma\wedge \tilde\diff\xi_\Delta\wedge
\tilde\diff\xi_\Lambda\rangle\with
\tilde\diff = \diff x^{\hat\imath}\frac{\pa}{\pa x^{\hat\imath}}
\end{equation}
are the constant components of the unique Lorentz-invariant three-form on $G$ whose values are written down in \cite{16}.
Thus, adding the functional (\ref{43b}) with a proper factor $\vk$ to the action (\ref{9}), we will get the Green-Schwarz
superstring action in the adiabatic limit $\ve\to 0$.

\medskip

\noindent {\bf $B$-field.} In string theory the action (\ref{37}) is often extended by adding the $B$-field term. This term
can be obtained from the topological Yang-Mills term
\begin{equation}\label{48}
\int^{}_{M^4} \langle \Fcal\wedge\Fcal\rangle = \sfrac14\,\int_{M^4} \diff^4x\, \ve^{\mu\nu\lambda\sigma}
\langle\Fcal_{\mu\nu}\, \Fcal_{\lambda\sigma}\rangle
\end{equation}
which in the adiabatic limit $\ve\to 0$ becomes
\begin{equation}\label{49}
\int^{}_{M^4} \diff^4x\, \ve^{ab} \ve^{ij}\langle\Fcal_{ai}\, \Fcal_{bj}\rangle =\int_{\Si_2} \diff x^1\diff x^2 \,
\ve^{cd}\, B_{\a\b}\, \pa_c X^\a \pa_d X^\b\ ,
\end{equation}
where
\begin{equation}\label{50}
 B_{\a\b}= \int_{H^2} \diff x^3\diff x^4 \,\ve^{ij}\langle\xi_{\a i}\, \xi_{\b j}\rangle
\end{equation}
are components of the two-form $\mathbb{B}=(B_{\a\b})$ on the moduli space $G$.

\medskip

\noindent {\bf AdS$_5\times S^5$ background.} We considered supergroup gauge theory on $\Si_2\times H^2$ with the structure
supergroup $G{=}$SUSY$(N=2)/$SO(9,1) and its reduction to the Green-Schwarz superstring theory with the target space $G$ in
the low-energy limit when $H^2$ shrinks to a point. Following the discussion in Section 2 and 3, one can see that everything
is similar to other supergroups and, in particular, for $G{=}$PSU(2,2$|$4) and the supercoset
PSU(2,2$|$4)/SO(4,1)$\times$SO(5). Namely, instead of the Lie algebra (\ref{1}) with the metric (\ref{2}) one should consider
the algebra $psu(2,2|4)$ whose commutation relations can be taken e.g. from~\cite{4t}, where type IIB GS strings moving in
AdS$_5\times S^5$ background were considered. Then the discussion of Section 2 and 3 can be repeated until the formula
(\ref{25}). Here one can take $h_\vp$ as an element of the supercoset PSU(2,2$|$4)/SO(4,1)$\times$SO(5) which will be the
moduli space. Then in the adiabatic limit $\ve\to 0$ one will get
\begin{equation}\label{51}
\Pi^{\Delta}_a=\pa_aX^M\, L_M^{\Delta}\ ,
\end{equation}
where $L_M^{\Delta}$ are components of the one-forms $L^{\Delta}= L_M^{\Delta}\diff X^M$ on the coset
PSU(2,2$|$4)/SO(4,1)$\times$SO(5) and the index $\Delta$ runs the coset parts of the generators of $psu(2,2|4)$~\cite{4t}. The
explicit form of the superstring action in terms of (\ref{51}) (both kinetic and WZ term) can be found in~\cite{4t}. Thus,
type IIB superstring action on AdS$_5\times S^5$ background can also be embedded into a supergroup gauge theory as the
low-energy limit.

\section {Concluding remarks}

We have introduced a Yang-Mills-Chern-Simons model  whose action functional in the low-energy limit reduces to the
Green-Schwarz superstring action. It was shown that $B$-field and Wess-Zumino-type terms in string theory appear from the
Yang-Mills topological terms  (\ref{48}) and (\ref{43b}), respectively. Combining these results with the results for bosonic
string \cite{7}, one can show that heterotic string theory can also be embedded into Yang-Mills theory as a subsector of
low-energy states. In fact, the described correspondence is a new kind of gauge/string duality. Thus, all five superstring
theories can be described in a unified manner via infrared limit of Yang-Mills-Chern-Simons theory. Such supergroup gauge
theories almost never been studied in the literature (see a discussion in~\cite{MW, 12t}).

The fibres of the considered Yang-Mills bundles are the supermanifolds $G$ with Minkowski space $\R^{9,1}$ as the bosonic part
or the supercosets PSU(2,2$|$4)/SO(4,1)$\times$SO(5) with AdS$_5\times S^5$ as the bosonic part. Considering quantum
Yang-Mills-Chern-Simons theory for small perturbations of Minkowski metric (or AdS$_5\times S^5$-metric), one gets
perturbative description of quantum gravity which in the adiabatic limit will reduce to the stringy description. Put
differently, embedding superstring theory on $\Si_2$ into Yang-Mills theory on $\Si_2\times H^2$ allows one to raise the
description of quantum gravity from stringy level to the Yang-Mills level. Then the gravitation will be described via
Yang-Mills fields for the diffeomorphism structure group at least perturbatively.

Last but not least, quantum Yang-Mills theory is developed incomparably better than quantum string theory. Calculations in
quantum Yang-Mills theory are much easier than in string theory.  Of course, to develop quantum Yang-Mills theory with gauge
supergroup $G$ is not a technically simple task. Some first steps have been done in~\cite{12t}. Anyway, studying such
Yang-Mills models, including the Chern-Simons term, can improve understanding of many questions in string theories, e.g. the
geometric sense of kappa symmetry and questions related with various dualities and AdS/CFT correspondence, as was argued
in~\cite{12t}.

\medskip

\noindent {\bf Acknowledgements}

\noindent This work was partially supported by the Deutsche Forschungsgemeinschaft grant LE 838/13.

%\newpage

\end{document}